\documentclass[12pt]{iopart}

\usepackage{iopams}
\usepackage{graphicx}
\begin{document}

\title[Probabilistic interconnection promotes cooperation]{Probabilistic interconnection between interdependent networks promotes cooperation in the public goods game}

\author{Baokui Wang$^1$, Xiaojie Chen$^2$ and Long Wang$^3$}

\address{
$^1$ Center for Complex Systems - Xidian University, Xi'an 710071, China\\
$^2$ Evolution and Ecology Program, International Institute for Applied Systems Analysis (IIASA), Schlossplatz 1, A-2361 Laxenburg, Austria\\
$^3$ Center for Systems and Control, State Key Laboratory for Turbulence and Complex Systems - Peking University, Beijing 100871, China
}

\ead{
\mailto{wangbaokui660@gmail.com},
\mailto{chenx@iiasa.ac.at} and
\mailto{longwang@pku.edu.cn}
}

\begin{abstract}
Most previous works study the evolution of cooperation in a structured population by commonly employing an isolated single network.
However, realistic systems are composed of many interdependent networks coupled with each other, rather than the isolated single one.
In this paper, we consider a system including two interacting networks with the same size, entangled with each other by the introduction of
probabilistic interconnections.
We introduce the public goods game into such system, and study how the probabilistic interconnection influences the evolution of cooperation of the whole system and the coupling effect between two layers of interdependent networks.
Simulation results show that there exists an intermediate region of interconnection probability leading to the maximum cooperation level in the whole system.
Interestingly, we find that at the optimal interconnection probability the fraction of internal links between cooperators in two layers is maximal.
Also, even if initially there are no cooperators in one layer of interdependent networks, cooperation can still be promoted by probabilistic interconnection, and the cooperation levels in both layers can more easily reach an agreement at the intermediate interconnection probability.
Our results may be helpful in understanding the cooperative behavior in some realistic interdependent networks and thus highlight the importance of probabilistic interconnection on the evolution of cooperation.
\end{abstract}


\maketitle

\section{Introduction}
The emergence of cooperation among selfish individuals in real world is still a challenging problem in social and biological systems~\cite{Axelrod1981}.
Evolutionary game theory has provided a uniform framework to improve our understanding of the emergence and sustenance of cooperation among unrelated individuals~\cite{Szabo2007,Roca2009,Perc2010}.
The public goods game (PGG), as one of the most famous paradigms, is often used for discussing the conflict between individuals and common interests~\cite{Brandt2003}.

In a typical PGG, cooperators (C) contribute an amount $c$ to the public good and defectors (D) do not contribute. The total contribution is multiplied by an enhancement factor $r$ ($r>1$) and then equally distributed by all the members in the group. Therefore, Ds obtain the same benefit of Cs with no cost, which confronts the individuals with the temptation to defect by taking advantage of the public good without contribution. Correspondingly, the Tragedy of the Commons is induced.
In order to elucidate why cooperators thrive under the exploitation of defectors in PGG, many mechanisms have been proposed, such as voluntary participation~\cite{Hauert2002,Szabo2002},
social diversity~\cite{Santos2008,Gao2010,Shi2010,Santos2012},
punishment~\cite{Helbing2010,Rand2011,Szolnoki2011,Sigmund2010},
migration~\cite{Cardillo2012,Wu2012},
reward~\cite{Szolnoki2010,Forsyth2011},
coordinated investments~\cite{Vukov2011},
the Matthew effect~\cite{Perc2011},
adaptive and bounded investment returns~\cite{Chen2012},
and conditional strategies~\cite{Szolnoki2012}.
In particular, the evolution of cooperation in the PGG on complex networks receives more attention recently.
It is shown that the mesoscale structure plays an important role in promoting the evolution of cooperation on complex networks~\cite{GG2011EPL,GG2011Chaos}.

It is worth pointing out that all the previous works enumerated above are based on the limited case of an isolated single complex network, e.g., square lattice, regular ring, or scale-free network.
However, empirical evidences show that real world systems are not isolated or disintegrated, but constructed by lots of interdependent networks, which connect and influence one another directly or indirectly~\cite{Gao2012}.
In other words, the real world in which we are living is a huge network of networks~\cite{Parshani2010,Rinaldi2001,Newman2005,Brummitt2012,Son2012}.
Especially in the social environment, there exists a diverse kinds of related social networks, such as the collaboration network and the social network.
It is known that, because of the social attribute of human beings, these two networks are directly linked together to some degree for scientists~\cite{Newman2001}.
Each individual on one layer of networks may probably link with the corresponding individual on the other one.
In this situation, the two layers may entangle with each other to different degrees.
In particular, this relationship can be described by the probabilistic interconnection between them.
Thus, it is interesting to study the social dynamics on the realistic system constructed by at least two layers of interdependent networks with probabilistic interconnection, and far less attention has been paid on the evolution of cooperation in this type of system.
Moreover, we would like to distinguish the interdependent networks from community structure and hierarchical network by network structure and social implication.
From the network structure, in a community structure, the connections are tight in communities but sparse between them~\cite{Girvan2002,Newman2012}.
In a hierarchical network, small groups of nodes organize in a hierarchical manner into increasingly large groups, while maintaining a scale-free topology~\cite{Ravasz2003}.
However, in interdependent networks, the interacting layers are connected through intra-layer links, which are independent of the interacting networks and can be tight or sparse in different situations.
The probabilistic interconnection between the interconnected layers determines the coupling effect between them~\cite{Buldyrev2010}.
From the social implication, the communities in a community structure probably share common properties and play similar roles~\cite{Fortunato2010}.
They can interact with the others to complete one overall functionality.
While in a hierarchical network, the individual in the center of bigger cluster means being at higher level in the hierarchy~\cite{Vukov2005}.
However, the layers of interdependent networks may have different properties or roles.
They interact with each other, influence each other and complement each other~\cite{Gao2011}.
Remarkably, recently Wang \etal studied the impact of biased utility functions on interdependent networks which were connected by utility functions~\cite{Wang2012}.
They showed that the benefits of enhanced public cooperation on the interdependent networks are as biased as the utility functions.
They emphasized that the positive effect of biased utility functions is due to the suppressed feedbacks of individual success, which leads to a spontaneous separation of characteristic time scales of the evolutionary process on interdependent networks.
However, we would like to point out that interacting layers in their work are only coupled by the utility functions, and indeed they may be also connected by probabilistic interconnection.

\begin{figure}
\centering
\includegraphics[width=6cm]{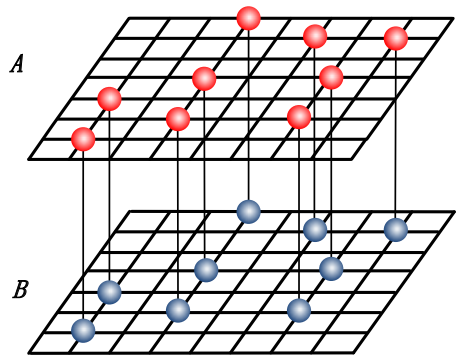}
\caption{Interdependent networks composed of two layers $A$ (denoted by red nodes) and $B$ (denoted by blue nodes).
Individuals are arranged on two square lattices of the same population size.
Under the probabilistic interconnection, a fraction of nodes in layer $A$ are connected with the corresponding nodes in layer $B$.
Here, a node from one layer can connect at most one node from the other layer.
}
\label{fig.1}
\end{figure}
In this paper, we develop a simple model of interdependent networks including two interacting layers connected through probabilistic interconnection to study the evolution of cooperation on them and the coupling effect between them.
For simplicity, we employ two identical spatial structures $A$ and $B$ of the same size.
Moreover, we assume that each node in $A$ connects the corresponding node in $B$ with probability $p$.
Here the interconnection probability $p$, which controls the number of links between $A$ and $B$, represents the integration of two interconnected layers.
For $p=0$, there are no internal links between the two layers. In other words, $A$ and $B$ are totally separated.
In the opposite limit, that is, $p=1$, all the nodes in $A$ and $B$ are completely connected in order.
For $0<p<1$, the actual number of internal links between $A$ and $B$ is subject to a binomial distribution.
We find that there exists an optimal intermediate region of the interconnection probability $p$ maximizing the cooperation level in the whole system.
Surprisingly, even if initially there are no cooperators in one layer of interdependent networks, by means of the coupling effect between them, the cooperation level on both layers can more easily reach an agreement at an intermediate interconnection probability.

\section{Model}

We investigate the evolutionary PGG on two interacting layers $A$ and $B$, as shown in figure 1. In order to focus explicitly on the impact of the probabilistic interconnection and easily compare our results with previous works, the two layers of interdependent networks we employed are both $L \times L$ square lattices with periodic boundary conditions and von Neumann neighborhoods. They have the same population size, $N_A = N_B$.
Each node in $A$ connects the corresponding node in B with probability $p$.
Meanwhile, we assume that a node from one layer connects to no more than one node from the other layer.

Initially, individuals in the system are designated either as a cooperator or a defector with equal probability.
Depending on the local links between individuals and the internal links between two interacting layers, individuals not only engage in five local PGGs which are centered on himself and the nearest neighbors on the same layer, may but also engage in one long-range PGG which is centered on the corresponding node on the other layer.
In addition, cooperators contribute $c=1$ to every PGG involved, and defectors contribute nothing. The total contribution is subsequently multiplied by the enhancement factor $r$, and then shared equally by all of the group members irrespective of their strategies.
The results obtained below are from the renormalized PGG enhancement factor $\eta=r/(M+1+p)$~\cite{Hauert2002,Gao2010,Vukov2011,Wu2009}, where $M=4$ is the number of local nearest neighbors of focal individuals.
The payoff of defectors engaging in one PGG is $P_d=r\cdot{n_c}/(M+1+p)={\eta}\cdot{n_c}$, and the corresponding payoff of cooperators is then $P_c=P_d-1$, where $n_c$ is the number of cooperators in the group.
After engaging in all the groups, player $x$ is allowed to learn from a randomly selected neighbor $y$ including the long-range one.
To be specific, player $x$ adopts the strategy of the random neighbor $y$ with a probability determined by the difference of their payoffs
\begin{equation}
\label{eq.1}
W_{(x{\leftarrow}y)}=\frac{1}{1+\exp[(P_{x}-P_{y})/{\kappa}]},
\end{equation}
where $\kappa$ denotes the noise effect in the strategy adoption process~\cite{Szabo1998}. Following previous study~\cite{Szolnoki2009a}, we simply set $\kappa=0.5$ in this work, and mainly focus on the impact of probabilistic interconnection.

\section{Simulation and analysis}

\begin{figure}
\centering
\includegraphics[width=13cm]{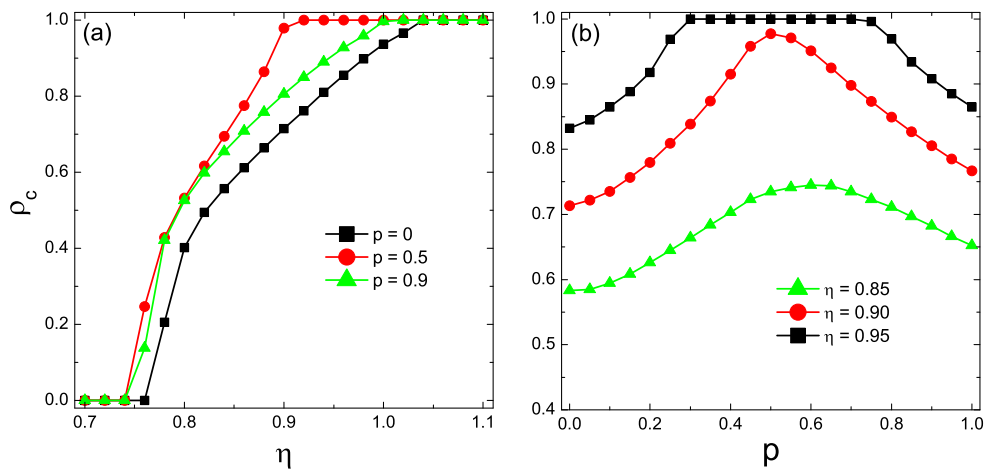}
\caption{The frequencies of cooperators among interdependent networks as a function of normalized enhancement factor $\eta$ with different values of $p$ in (a) and as a function of interconnection probability $p$ with different values of $\eta$ in (b).}
\label{fig.2}
\end{figure}
In the following, we show the simulation results carried out on the interdependent networks composed of two interacting layers both of size $100\times100$.
It is known that network size can strongly influence the dynamics of such system~\cite{Szolnoki2009b}.
Thus, we have also conformed our model in different systems of larger network size, e.g., $200\times200$.
The simulation indicates that our results are robust in larger systems.
Initially, cooperators and defectors are randomly distributed on both layers of interdependent networks with equal probability.
We define that $\rho_{ac}$ denotes the frequency of cooperators of $A$, and $\rho_{bc}$ the frequency of cooperators of $B$ correspondingly.
Meanwhile, we define that $\rho_{c}$ denotes the frequency of cooperators in the whole system.
In our work, we adopt the synchronous Monte Carlo simulation procedure to update the strategies of players. Unless otherwise stated, all the simulation results shown below are required up to $3\times10^{4}$ generations and then sampled by another $10^{3}$ generations. The results are averaged over $30$ realizations of different initial conditions.

We first plot the frequency of cooperators in the whole system as a function of the renormalized enhancement factor $\eta$ for different values of $p$ in figure 2(a).
For $p=0.5$, we find that the cooperation level is enhanced obviously on a large scale of $\eta$ and the full cooperation state is achieved at $\eta=0.92$.
While $p=0$, we note that the full cooperation state is achieved at $\eta=1.04$, which falls behind $p=0.5$.
For $p=0.9$, at the same values of $\eta$, the cooperation level is higher than that of $p=0$ but lower than that of $p=0.5$, and the full cooperation state is achieved at $\eta=0.98$.
We note that, as shown in figure 2(a), the probabilistic interconnection between two interacting layers can significantly influence the evolution of cooperation in the whole system.
In order to study the role of probabilistic interconnection in promoting cooperation intensively, we investigate the frequencies of cooperators as a function of $p$ with different $\eta$ in figure 2(b).
We observe that, for each fixed $\eta$, there exists an optimal intermediate region of $p$ maximizing the cooperation level in the whole system.
The probabilistic interconnection promoting cooperation resembles an interesting resonancelike phenomenon reflected by the optimal cooperation level at the intermediate interconnection probability.
Additionally, we find that, for each fixed $\eta$, the cooperation level in the whole system $\rho_{c}$ at $p=1.0$ is larger than that at $p=0$.
In addition, we would like to point out that it is difficult to use theoretical analysis, e.g. pair-approximation method, to investigate the cooperation level in this system.
However, these simulation results presented in figure 2 clearly evidence that the introduction of probabilistic interconnection between interdependent networks significantly influences the evolution of cooperation on them and there exists some intermediate values of $p$ maximizing the cooperation level in the whole system.

\begin{figure}
\centering
\includegraphics[width=8cm]{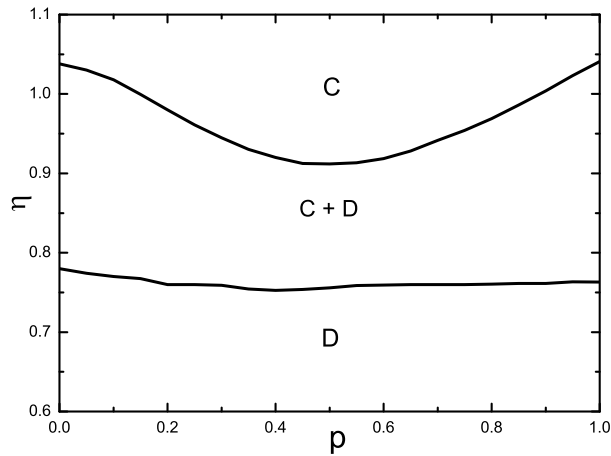}
\caption{The complete $p-\eta$ phase diagram for PGG on the interdependent networks. The upper (lower) boundary is the extinction thresholds of cooperators (defectors) correspondingly.}
\label{fig.3}
\end{figure}
In figure 3, we draw the full $p-\eta$ phase diagram for the evolution of cooperation on the interdependent networks.
It is worth noting that, for each value of $p$, there exists a lower critical value and an upper critical value for $\eta$ respectively. Below the lower critical value, defectors dominate the whole population; while above the upper critical value, cooperators dominate the whole population.
We observe that for intermediate $p$, both the upper and lower boundaries attain their minimum values, where the coupling effect between interdependent networks is the strongest.
Clearly, the upper critical value of $\eta$ first monotonously decreases until reaching the minimal value at about $p=0.5$, then increases with increasing $p$.
This result further indicates that the impact of the effective interaction topology induced by the probabilistic interconnection reaches the strongest at intermediate $p$.

\begin{figure}
\centering
\includegraphics[width=8cm]{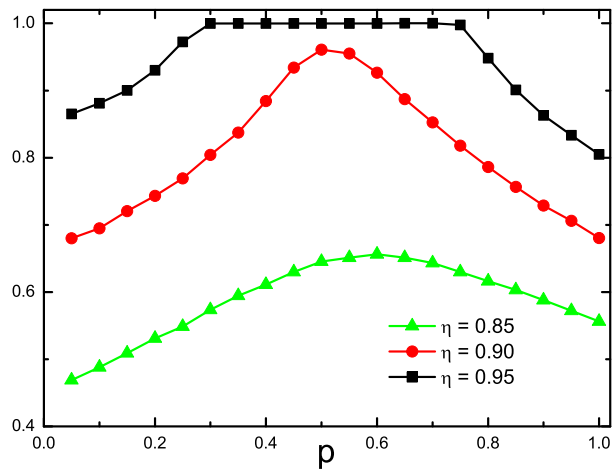}
\caption{The frequencies of $CC$ links between two interacting layers as a function of the probabilistic interconnection $p$ with different $\eta$.}
\label{fig.4}
\end{figure}
In order to intensively investigate the coupling effect induced by probabilistic interconnection, we calculate the frequencies of $CC$ links between the interacting layers at equilibrium with different $\eta$ in figure 4.
Interestingly, we find that the frequencies of $CC$ links with different $\eta$ between two layers are pretty similar with that of cooperators in the whole system shown in figure 2(b) for corresponding values of $\eta$.
This phenomenon indicates that the coupling effect due to the probabilistic interconnection between interdependent networks fundamentally influences the cooperation level in the whole system.
To describe the strategy relationship between two interacting layers, especially to describe the assortment of $CC$ links between them quantitatively, we employ the correlation coefficient between individuals~\cite{Robert1994}
\begin{equation}
\label{eq.2}
r_c=\frac{E({S_i}{S_j})-E(S_i)E(S_j)}{\sqrt{E(S_i^{2}-E(S_i)^{2})}\sqrt{E(S_j^{2}-E(S_j)^{2})}},
\end{equation}
where $S_i$ and $S_j$ are the strategies of a random pair of connected individuals $i$ and $j$. If $i$ is a $C$, $S_i=1$. Otherwise, $S_i=0$. $E(\cdot)$ is the expected value of corresponding strategy.
In an infinite system, we approximately consider a simplified correlation coefficient $r_c=(\rho_{CC}-\rho_{c}^{2})/(\rho_{c}-\rho_{c}^{2})$,
where $\rho_{CC}$ is the fraction of $CC$ links between two layers.
We calculate that $r_c>0$ with different $\eta$, when $0<\rho_c<1$.
It means that there exists strong positive strategy relationship between two layers.
Thus, the probabilistic interconnection which fundamentally determines the evolution of cooperation in spatial public goods game strengthens the coupling effect between interdependent networks.

\begin{figure}
\centering
\includegraphics[width=11cm]{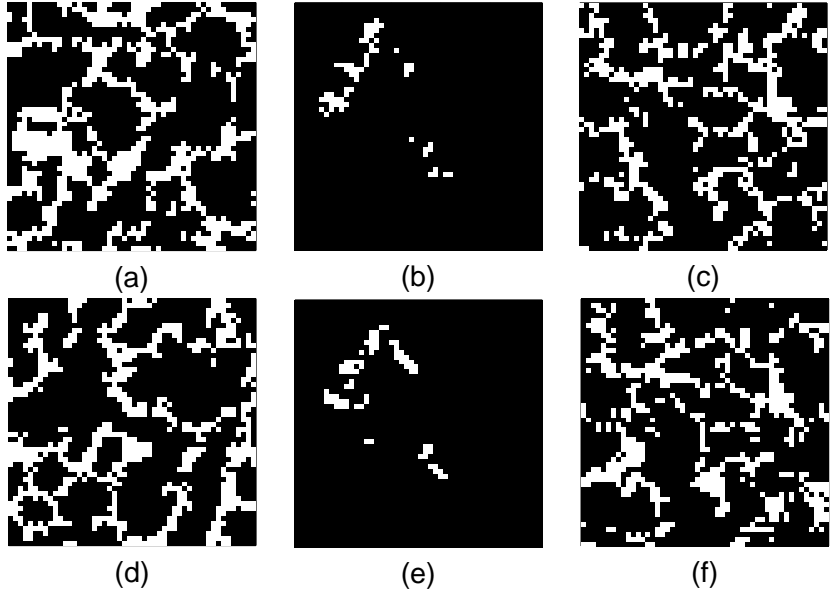}
\caption{Snapshots of the typical distributions of cooperators (black) and defectors (white) on $A$ (\textbf{a-c}) and $B$ (\textbf{d-f}) obtained by $\eta=0.9$ and different values of $p$.
These snapshots are a $50\times50$ portion of the full $100\times100$ lattices.
(a) $p=0$ ($\rho_{ac}=0.7024$), (b) $p=0.5$ ($\rho_{ac}=0.9744$), (c) $p=1.0$ ($\rho_{ac}=0.768$),
(d) $p=0$ ($\rho_{bc}=0.736$), (e) $p=0.5$ ($\rho_{bc}=0.9704$) and (f) $p=1.0$ ($\rho_{bc}=0.7648$).}
\label{fig.5}
\end{figure}
To understand this coupling effect intuitively, in the following we present some snapshots of the numerical simulation results of both $A$ and $B$ at stationary state for fixed $\eta=0.9$ and three different values of $p$, as shown in figure 5.
It can be observed that, for intermediate $p=0.5$, cooperators can dominate defectors in both $A$ and $B$.
Whereas for small $p=0$ or large $p=1.0$, cooperators can only coexist with defectors in the long run.
These snapshots demonstrate that, at intermediate $p$, the cooperation level in $A$ and $B$ is respectively higher than ones for other values of $p$, thus the fraction of cooperators in the whole population is highest at these intermediate $p$ values.
Moreover, for different values of $p$, the cooperation level in $A$ is similar to that in $B$.
And similar spatial patterns simultaneously display in $A$ and $B$ especially for intermediate $p$, which indicates the strong coupling effect exhibit between these two interacting layers of interdependent networks.

\begin{figure}
\centering
\includegraphics[width=11cm]{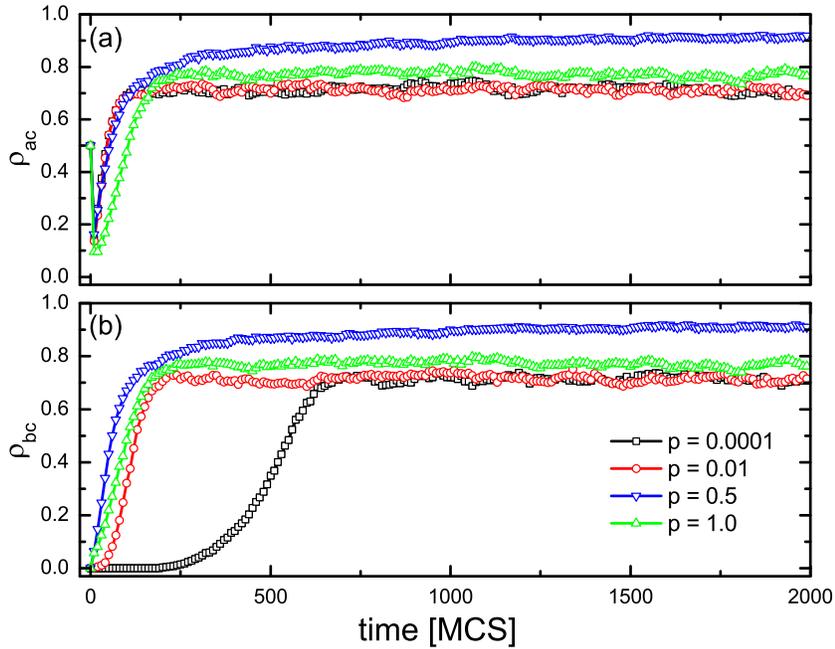}
\caption{Simultaneous time evolution of cooperation level on $A$ (${\rho}_{ac}$) and $B$ (${\rho}_{bc}$) for $\eta=0.9$.
Initially, cooperators and defectors are randomly distributed on $A$ with equal probability, but the individuals on $B$ are all defectors.}
\label{fig.6}
\end{figure}
Finally, to further explore the coupling effect induced by probabilistic interconnection between $A$ and $B$ on the evolution of cooperation, in figure 6 we study the time evolution of fraction of cooperators in $A$ and $B$ respectively by assuming that two different initial conditions.
Initially, cooperators and defectors are randomly distributed on $A$, while only defectors are distributed on $B$.
In figure 6(a), we find that the cooperation level for different values of $p$ first decreases sharply and then rapidly increases to the equilibrium state.
The fraction of cooperators in $B$ monotonously increases with time for different values of $p$, even for very small $p=0.0001$.
Interestingly, for each value of $p$, the final fraction of cooperators in $B$ at equilibrium is similar to the value in $A$.
Remarkably, the fractions of cooperators in $A$ and $B$ can easily reach an agreement for intermediate interconnection probability $p=0.5$.
Also, the final cooperation levels on both layers for intermediate $p$ are higher than that of the other values of $p$.
These results imply that the strongest coupling effect between the two interacting layers emerges at intermediate interconnection probability, which synchronously promotes the evolution of cooperation in the whole system.

\section{Discussion and conclusion}
Let us further discuss the differences between our model and some relevant previous works~\cite{Vukov2005,Wang2012,Chen2007,Wang2011,GG2012SR}.
In ref.~\cite{Chen2007}, Chen \etal studied the evolutionary Prisoner's Dilemma game on a community structure network which exhibits scale-free property. They showed that the cooperation level decreases with the increment of the average degree and reducing inter-community links can promote cooperation when keeping the total links unchanged.
In ref.~\cite{Wang2011}, Wang \etal analytically studied the evolution of cooperation in multilevel public goods games with community structures. They demonstrated that cooperation and punishment are more abundant than defection in the case of sufficiently large community size and number with different imitation strength between communities.
In ref.~\cite{Vukov2005}, Vukov and Szab\'{o} studied how the cooperation level is affected by the number of hierarchical levels and by the temptation to defect.
They showed that the highest frequency of cooperation can be observed at the top level if the number of hierarchical levels is low, and for larger number of hierarchical levels, the highest cooperation level occurs in the middle layers.
However, as we described above, these works are studied in an isolated single network with community or hierarchical structure which does not fully investigate the complex nature of the real world composed of interdependent networks~\cite{Gao2012,Parshani2010,Brummitt2012,Buldyrev2010}.
In our study, the interdependent networks are composed of two interacting layers and the internal links build a bridge between them.
They inspires individuals on different layers to interact with each other and then influence each other, thereby promoting the evolution of cooperation in the whole system and enhancing the coupling effects between interdependent networks.
It is worth pointing out that in our work, although the two interconnected lattices are linked directly by the probabilistic interconnection, it does not affect the nature of our research.
Even if the interaction process and learning process are detached, they can still influence the evolutionary dynamics of cooperation in the whole system eventually due to the coupling effects between interdependent networks.

Recently, G\'{o}mez-Garde\~{n}es \etal studied the evolutionary game dynamics on multiplex interdependent networks~\cite{GG2012SR}.
In their work, each individual plays with all the neighbors on different layers of networks and obtain the net payoff of all the payoffs collected in each of network layers by using a set of strategies.
They showed that the resilience of cooperation for extremely large values of the temptation to defect is enhanced by the multiplex structure.
Furthermore, this resilience is intrinsically related to a non-trivial organization of cooperation across the network layers, thus providing
a new way out for cooperation to survive in structured populations.
However, the coupling factors among layers of networks are not considered in their work.
Whereas in our study, the introduction of probabilistic interconnection controlled by parameter $p$ communicates one layer to another, introducing the coupling effects.
Moreover, the interconnection probability $p$ indicates the integration of the two interacting layers of interdependent networks.
Unlike the previous works, the internal links in our work significantly influence the dynamical behavior in the whole system by the coupling effects.
There resembles an interesting resonancelike phenomenon~\cite{Perc2006NJP,Ren2007} reflected by the optimal cooperation level at the intermediate interconnection probability.
Also, introducing appropriate probabilistic interconnection between interdependent networks can significantly influence the formation process of clusters which determines the evolution of cooperation in the whole system to a great extent.
Interestingly, the patterns of clusters in both networks are very similar even by the approximate locations of cooperators and defectors, which presents an even richer resonancelike behavior.
Furthermore, comparing the results of $p=0$ with $p=1$, we find that the average level of cooperation when $p=1$ is always larger than that of $p=0$, thus enforcing the positive role of probabilistic interconnection between two interacting layers.
Therefore, our study further complements the investigation about coupling effect between interdependent networks in the framework of evolutionary graph theory, and enriches the knowledge of evolutionary dynamics in the PGG.

In summary, we have studied the evolution of cooperation on two interacting layers of interdependent networks coupled by probabilistic interconnection.
We have shown that the introduction of probabilistic interconnection provides a new way of understanding the emergence and maintenance of cooperation among selfish individuals in sizable groups.
We find that there exists an optimal intermediate region of $p$ maximizing the cooperation level in the whole system.
Importantly, we clearly evidence that the introduction of probabilistic interconnection between interdependent networks strengthens the coupling effect between them.
Therefore, the probabilistic interconnection between interacting layers just like a bridge opening the way for the neighboring networks to interact with each other.
By means of this simple model, we would like to reveal the internal mechanisms about how the evolution of cooperation thrives in the real world constructed by interdependent networks.
Although this model is simple and does not include every kind of circumstances existing, we hope this beneficial attempt can highlight the way of exploring the internal mechanisms in promoting the evolution of cooperation on interdependent networks which is closer to reality.

\ack
We gratefully thank J. Gao, Y. Liu, T. Wu and X. Wang for useful discussion and comments. This work was supported by the National 973 Program (Grant No. 2012CB821203), the National Natural Science Foundation of China (Grants No. 61020106005, 11161011, 10972002 and 61104212) and the Fundamental Research Funds for the Central Universities (Grants No. K50511040005 and K50510040003).

\end{document}